\documentstyle[12pt,epsfig]{article}

\newcommand{\bea}{\begin{eqnarray}}
\newcommand{\eea}{\end{eqnarray}}
\newcommand{\beq}{\begin{equation}}
\newcommand{\eeq}{\end{equation}}
\newcommand{\bay}{\begin{array}}
\newcommand{\eay}{\end{array}}
\begin{document}


\begin{flushright}
TECHNION-PH-2002-18\\
UCSD/PTH 02-10 \\
hep-ph/0205065 \\
May 2002 \\
\end{flushright}

\renewcommand{\thesection}{\Roman{section}}
\renewcommand{\thetable}{\Roman{table}}
\centerline{\bf PHOTON POLARIZATION IN RADIATIVE $B$ DECAYS}
\bigskip
\centerline{Michael Gronau}
\centerline{\it Physics Department, Technion -- Israel Institute of Technology}
\centerline{\it 32000 Haifa, Israel}
\medskip
\centerline{Dan Pirjol}
\centerline{\it Department of Physics, University of California at San Diego}
\centerline{\it 9500 Gilman Drive, La Jolla, CA 92093}
\bigskip

\begin{quote}
We study decay distributions in $B\to K\pi\pi\gamma$, combining contributions
from several overlapping  resonances in a $K\pi\pi$ mass range near 1400
MeV, $1^+~K_1(1400)$, $2^+~K^*_2(1430)$ and $1^-~K^*(1410)$. A method is 
proposed for using these distributions to determine 
a photon polarization parameter in the effective radiative weak Hamiltonian. 
This parameter is measured through an up-down asymmetry of the photon direction 
relative to the $K\pi\pi$ decay plane in the $K_{\rm res}$ frame.
We calculate a dominant up-down asymmetry of $0.33 \pm 0.05$ from the
$K_1(1400)$ resonance, which can be measured with about $10^8$ $B\bar B$ pairs,
thus providing a new test for the Standard Model
and a probe for some of its extensions.

\end{quote}

\section{Introduction}

Measurements of inclusive radiative $B$ meson decays 
$B \to X_s\gamma$ \cite{Bsg} provide an important test for the Standard Model 
(SM), and set stringent bounds on physics beyond the SM \cite{PG}. 
In addition to the rather well predicted inclusive branching ratio, which was 
studied extensively both experimentally and theoretically \cite{Misiak}, there 
is a unique feature of this process within the SM which drew only moderate 
theoretical attention and which has not yet been tested. 
Namely, the emitted photons are left-handed in radiative $B^-$ and $\bar B^0$ 
decays and are right-handed in $B^+$ and $B^0$ decays. In the SM the photon in 
$b \to s\gamma$ is predominantly left-handed, since the recoil $s$ quark which 
couples to a $W$ is left-chiral. This prediction of approximately
{\it maximal parity violation} holds in the SM to within a few percent, up 
to corrections of order $m_s/m_b$. It applies also in exclusive radiative 
decays when including long-distance effects \cite{GrinPi}. While measurements 
of the inclusive radiative decay rate agree with SM calculations,
no evidence exists for the helicity of the photons in inclusive and exclusive 
decays. 

In several extensions of the SM the photon in $b\to s\gamma$ 
acquires an appreciable right-handed component due to chirality flip
along a heavy fermion line in the electroweak loop process. Two well-known 
examples of such extensions are the left right symmetric (LR) model and the 
unconstrained Minimal Supersymmetric Standard Model (MSSM). In the LR model 
chirality flip along the $t$ quark line in the loop involves $W_L-W_R$ 
mixing \cite{LRM}, while in the MSSM
a chirality flip along the gluino line in the loop involves left-right squark 
mixing \cite{MSSM}. In both types of models it was found that, in certain 
allowed regions of the parameter space, the photons emitted in $b\to s\gamma$ 
can be largely right-handed polarized, without affecting the SM prediction for 
the inclusive radiative decay rate. This situation calls for an independent 
measurement of the photon helicity, which therefore becomes of immediate 
interest.

Several ways were suggested in the past five years to look for signals of 
physics beyond the SM through photon helicity effects in $B \to X_s \gamma$.
To set the stage for the present proposal, and in order to appreciate the 
immediate potential of applying this new idea at currently operating $B$ 
factories, while other methods require higher luminosities or new experimental 
facilities, let us recall all previously proposed methods \cite{HF9}. 

In  the first suggested method \cite{AGS} the photon helicity is 
probed through mixing-induced CP asymmetries. The time-dependent asymmetry of 
$B^0(t) \to X^{CP}_{s(d)}\gamma$, where $X^{CP}_s = K^{*0} \to K_S\pi^0$ or 
$X^{CP}_d = \rho^{0} \to \pi^+\pi^-$, follows from interference between $B^0$ 
and $\bar B^0$ decay amplitudes into a common state of definite photon 
polarization. The asymmetry is proportional to the ratio of 
right-to-left polarization amplitudes $A_R/A_L$, for small values of this ratio 
(a few percent in the SM), and may reach a maximum value of order one in 
extensions of the SM. For a time-dependent measurement, one must measure the 
distance of the $B$ decay point away from its production. It is hard to trace a 
$K^{*0}$ decay back to its point of production, since in $K^{*0} \to K_S \pi^0$ 
the $K_S$ decays after travelling some distance. This is not the case for $B^0 \to 
\rho^0\gamma$, where the $\rho^0$ decays promptly to $\pi^+\pi^-$, allowing 
thereby a time measurement. However, $B^0 \to \rho\gamma$ is suppressed by
a factor $|V_{td}/V_{ts}|^2$ relative to ${\cal B}(B^0 \to K^*\gamma) \sim 
4\times 10^{-5}$ \cite{K*gam}, and one expects its branching ratio to be only 
about $2\times 10^{-6}$.  CP asymmetries at a level of a few percent, as 
expected in the SM, require an order of $10^9$ $B$ mesons. A smaller number 
of order $10^8~B$'s, as envisaged in near future 
experiments at $B$ factories, can provide constraints 
on possible large asymmetries, namely on large values of $A_R/A_L$, and would 
thereby improve bounds on certain parameters of the new physics models.

In a second scheme one studies angular distributions in
$B\to \gamma (\to e^+e^-) K^* (\to K\pi)$, where the photon can be 
virtual \cite{Kim} or real, converting in the beam pipe to an electron-positron 
pair \cite{GrPi}. The correlation between the $e^+e^-$ and the $K^*\to K\pi$ planes 
is sensitive to the photon polarization. The distribution in 
the angle between the $K\pi$ and $e^+e^-$ planes is isotropic for purely 
circular polarization, and the angular distribution is sensitive to 
interference between left and right polarization. Namely, the deviation from 
isotropy involves (again) a parameter $A_R/A_L$ measuring the mixture of left 
and right polarizations. 
One expects ${\cal B}(B\to K^*e^+e^-) \sim (1-2)\times 10^{-6}$ \cite{K*ee}.
Therefore, the number of $B$'s required here to measure a 
photon polarization effect, in the SM or in the presence of new physics, is 
comparable to the corresponding number required for the previous method. 

In a third method using $\Lambda_b$ decays \cite{MaRe}, $\Lambda_b 
\to \Lambda\gamma \to p\pi\gamma$, one measures directly the photon polarization. 
The forward-backward asymmetry of the proton with respect to the $\Lambda_b$ 
in the $\Lambda$ rest-frame is proportional to the photon polarization. 
Using polarized $\Lambda_b$'s \cite{HK}, one can also measure the 
forward-backward asymmetry of the $\Lambda$ momentum with respect to the 
$\Lambda_b$ boost axis. This asymmetry is proportional to the product of the 
$\Lambda_b$ and photon polarizations. This last scheme can only be applied in 
extremely high luminosity $e^+e^-$ $Z$ factories. We note that the two methods
based on $\Lambda_b$ decays
measure directly the photon polarization, whereas the other two types of 
measurements using $B$ decays are sensitive to interference between amplitudes 
involving photons with left and right-handed polarization. All methods can 
probe deviations from approximately
pure left-handedness as predicted in the SM.

In the present paper we wish to elaborate further on a method proposed very 
recently in a short Letter \cite{ggpr} (see also \cite{Weinstein}), which
measures directly a fundamental parameter in the effective radiative weak 
Hamiltonian describing the photon polarization. 
This method, based on radiative $B$ decays to excited kaons, makes use of 
angular correlations among the three-body decay products of the excited kaons. 
It was shown \cite{ggpr} that in decays 
$B^+\to (K^+_1(1400) \to K^0\pi^+ \pi^0)\gamma$ and $B^0\to (K^0_1(1400) \to 
K^+\pi^- \pi^0)\gamma$ the up-down asymmetry of the photon 
momentum with respect to the $K\pi\pi$ decay plane in the $K_{\rm res}$ frame
measures the photon 
polarization with a rather high efficiency. For approximately complete 
polarization, as expected in the SM, the asymmetry integrated over the entire 
Dalitz plot was calculated to be 0.34. Here we study this asymmetry in some 
more detail than in \cite{ggpr}. In 
particular, we calculate carefully theoretical uncertainties due to an
admixture of $S$ and $D$ waves in $K_1 \to K^*\pi$ and due to a 
possible small decay rate into $\rho K$. Assuming the radiative branching ratio 
into $K_1(1400)$ to be around but somewhat below $10^{-5}$ \cite{gK}, such an 
asymmetry can be measured at currently operating $B$ factories. 

The $K\pi\pi$ invariant mass region around 1400 MeV contains several kaon
resonances with different quantum numbers. We give further details for 
calculations of angular distributions corresponding to these resonance states.
Whereas the axial-vector state $K_1(1400)$ introduces a large 
up-down asymmetry, the other two states, a tensor $K^*_2(1430)$ and a vector 
$K^*_1(1400)$, lead  to a much smaller asymmetry and to no asymmetry, 
respectively. Separation or projection of these individual resonance 
contributions is therefore crucial, in order to achieve a high 
efficiency in measuring 
the photon polarization parameter in the effective radiative weak Hamiltonian.
It is shown that this parameter measures the photon polarization in decays to 
all individual $K$ resonances. 
Interference 
between the different resonances, which was disregarded in \cite{ggpr}, would 
introduce uncertainties in the measurement of the polarization parameter. 
Here we present details of a method, by which interference effects between 
overlapping resonances can be resolved, thereby providing a way of measuring 
the photon polarization parameter with only minimal model-dependence. 

In addition to the above-mentioned final states which involve a neutral pion, 
we also consider extensions of the method to two 
other resonance decay channels involving only charged pions, $K^+\pi^+\pi^-$ 
and $K^0\pi^+\pi^-$, not considered in \cite{ggpr}. We explain the sources of 
theoretical uncertainties in studying polarization effects in these decays.

The basic idea of the method is introduced in Sec.~II, where it is
shown in general that certain observables in $B\to (K_{\rm res}\to K\pi\pi)
\gamma$ decays are sensitive to the photon polarization. Several relevant
final states are listed to which this measurement can be applied.
These final states involve a kaon and two pions of specific charges. 
Sec.~III considers, for resonance states of specific quantum numbers, the 
general structure of the weak decay amplitude of $B\to (K_{\rm res}\to 
K\pi\pi)\gamma$ in terms of the photon polarization. 
It is shown that 
the polarizations corresponding to individual resonance states are all identical
to a polarization parametrer defined by Wilson coefficients in the effective 
weak Hamiltonian.
Strong decays of
several resonance states with different quantum numbers are studied
in Sec.~IV. Angular decay distributions of $B\to K\pi\pi\gamma$, sensitive 
to the photon polarization, are calculated in Sec.~V. A method is presented for 
determining the photon polarization parameter
from decay distributions, in spite of
involving contributions from several overlapping kaon resonances. 
Sec. VI presents numerical results for the expected 
up-down asymmetry parameters. Sec.~VII contains a discussion of the 
experimental feasibility of the method, followed by conclusions in Sec.~VIII.

\section{Why and which three body decays of $K_{\rm res}$?}

The first measured radiative $B$ decays \cite{K*gam} were exclusive decays 
into the first excited kaon resonance state, $B \to K^*(892)\gamma$, with 
branching ratios of about $4\times 10^{-5}$ .
Radiative decays into a higher excited resonance state, $K^*_2(1430)$, were  
observed more recently, both by the CLEO and Belle collaborations,
\bea\label{BR}
{\cal B}(B\to K_2^*(1430)\gamma) &=&
(1.66^{+0.59}_{-0.53}\pm 0.13)\times 10^{-5}\qquad\quad 
\mbox{(CLEO~\cite{CLEO})}~~,\nonumber\\
&=& ( 1.50^{+0.58+0.11}_{-0.53-0.13})\times 10^{-5}\qquad \qquad~~
\mbox{(Belle~\cite{Ishikawa})}~~.
\eea
In these experiments the $K^*(892)$ and $K^*_2(1430)$ resonance states were 
identified through their $K\pi$ decay channels. In both cases the 
corresponding $K\pi$ decay branching ratios are large, $(49.9\pm 1.2)\%$ in the 
case of $K^*_2$ \cite{PDG}. 

In the case of $K^*_2$, decay branching ratios into $K^*(892)\pi$ and $\rho K$ 
are also sizable, $(24.7\pm 1.5)\%$ and $(8.7\pm 0.8)\%$, respectively 
\cite{PDG}. These modes lead to $K\pi\pi$ final states. As we will argue below, 
in order to probe the helicity of the $K^*_2$ (or any other resonance), and 
thereby 
determine the photon polarization, one must study the resonance decays into 
final states involving at least three particles. First measurements of $B$
decays into a photon and three body hadronic final states, involving a 
$K\pi\pi$ invariant mass in the kaon resonance region, were reported recently 
by the Belle collaboration \cite{Ishikawa}. As will be explained below,
this experiment can be used to measure the photon polarization. 

Let us explain first the necessary conditions for a measurement of
the photon polarization through the recoil hadron distribution. We will
also consider the essential ingredients of the measured physical system which 
are necessary for a theoretically clean measurement, that is, a measurement
which involves a minimal amount of hadronic dependence. 

Since the photon helicity is odd under parity, and since
one only measures the momenta of the photon and of the final hadronic decay 
products, helicity information cannot be obtained from two body decays of the 
excited kaon. A hadronic quantity which is proportional to the photon helicity 
must be parity odd. The pseudoscalar quantity, which contains the smallest 
number of hadron momenta, is a triple product. Thus, one requires at least a
three body hadronic final state recoiling against the photon, in which one can 
form a parity-odd triple product $\vec p_{\gamma}\cdot (\vec
p_1\times \vec p_2)$ in the $K$ resonance rest frame. Here $\vec p_{\gamma}$ is 
the photon momentum, while
$\vec p_1$ and $\vec p_2$ are two of the final hadron momenta, all measured in
the recoiling hadron (a $K$-resonance) rest frame. Applying parity, the 
average value of the triple product has one sign for a left-handed photon and 
an opposite sign for a right-handed photon.

But here there seems to be a theoretical difficulty. A triple product correlation 
is also odd under time-reversal. Since time-reversal symmetry holds
in $K$-resonance decays, the decay amplitude must involve a nontrivial phase
due to final state interactions. Such a phase is usually suspected of being 
uncalculable and hard to measure. This strong phase originates from the 
interference of at least two amplitudes leading to a common three-body final 
state. Noting that the $K\pi\pi$ decay modes of excited resonance states
around 1400 MeV (to which we will draw our attention) are dominated by $K^*\pi$ 
and $\rho K$ channels \cite{PDG}, let us list the three kinds of interference 
which one may encounter:  

\begin{enumerate}
\item Interference between two intermediate $K^*\pi$ states with different 
charges, for instance $K^{*+}\pi^0$ and $K^{*0}\pi^+$, decaying to a common 
$K^0\pi^+\pi^0$ state.
These two amplitudes are related by isospin; consequently the strong phase
is calculable purely in terms of Breit-Wigner forms.
\item Interference between $K^*\pi$ and $\rho K$ amplitudes. In several cases 
these amplitudes can be related by SU(3), and SU(3) breaking can be obtained 
from measured decay branching ratios of excited kaons into $K^*\pi$ and 
$\rho K$. In some cases relative strong phases are extracted from resonance 
production experiments.
\item Interference between different partial waves into $K^*\pi$ or 
$\rho K$. In certain cases the ratio of these partial wave amplitudes and their 
relative phases were measured in resonance production experiments.
\end{enumerate}
Since we will consider a $K\pi\pi$ state with an invariant mass in a narrow
band ($\pm$~100 MeV) around 1400 MeV, we will neglect the direct 
nonresonant radiative $B\to K\pi\pi\gamma$ decay. The decay rate 
for this phase space restricted process is much smaller than the decay rate 
through an excited resonance state. A simple estimate of the nonresonant
contribution can be made by noting that 
the ratio of the resonant and nonresonant $K\pi\pi\gamma$ events with total
$K\pi\pi$ invariant mass within a region of width $\Gamma_{K_{\rm res}} \sim 200$ 
MeV is about $\sim \Gamma_{K_{\rm res}}/(M_B-(2M_K + M_\pi) \sim 4\%$.
We assumed here for simplicity equal total resonant and nonresonant branching
ratios, with a flat distribution in the $K\pi\pi$ invariant mass for the latter.

We conclude that the theoretically cleanest calculation of a $K\pi\pi\gamma$ 
decay amplitude, in which final state interaction phases can be computed 
most reliably from pure isospin considerations, corresponds to cases in which 
only the first kind of interference exists.
We will focus our attention on cases which are dominated by such interference, 
which permits a theoretically rather clean 
measurement of the photon polarization through decay distributions. 

In Sec.~IV we will study the decays of three $K$ resonances, $K_1(1400)$, 
$K^*(1410)$ and $K_2^*(1430)$, with quantum numbers $J^P = 1^+, 1^-$ and $2^+$,
respectively.  
The most general result for the decay amplitude involves relative strong phases 
between $K^*\pi$ and $\rho K$ amplitudes and between different partial waves. 
In order to have a measurement which can be cleanly interpreted in terms of 
the photon polarization parameter, these phase differences must be known, at 
least crudely. This is the case in the decays $K_1(1400) \to
K\pi\pi$, which are dominated by $K^*\pi$ intermediate states, and where 
some information is known both about the $S-D$ admixture in the $K^*\pi$ 
channel, and about the magnitude and phase of the smaller $K\rho$ amplitude.

Parametrizing resonance amplitudes in terms of Breit-Wigner forms,  
known to be a rather good approximation, yields a calculable strong phase.
As mentioned, the remaining strong phases can be estimated in some cases using 
arguments based on flavor SU(3) symmetry, or can be extracted from 
resonance production experiments.
In many respects, this method is similar to measuring the $\tau$ neutrino 
helicity in
$\tau \to a_1\nu_{\tau}$, where the corresponding phase-difference
is calculable in terms of the two interfering  $a_1 \to \rho\pi$ amplitudes
corresponding to two different $\rho\pi$ charge assignments 
\cite{Kuhn,Feindt,ARGUS}.

Let us list the channels through which excited kaons may decay into 
distinct charged $K\pi\pi$ states. 
\bea\label{chain1}
K_{\rm res}^{+}\to & &
\left\{
\begin{array}{c}
 K^{*+}\pi^0 \\
 K^{*0} \pi^+ \\
 \rho^+ K^0
\end{array}
\right\} \to K^0 \pi^+ \pi^0~~,\\
& &\label{chain2}
\left\{
\begin{array}{c}
 K^{*0} \pi^+ \\
 \rho^0 K^+
\end{array}
\right\} \to K^+ \pi^+ \pi^-~~,\\
\label{chain3}
K_{\rm res}^0\to & &
\left\{
\begin{array}{c}
 K^{*+}\pi^- \\
 K^{*0} \pi^0 \\
 \rho^- K^+
\end{array}
\right\} \to K^+ \pi^- \pi^0~~,\\
& &\label{chain4}
\left\{
\begin{array}{c}
 K^{*+} \pi^- \\
 \rho^0 K^0
\end{array}
\right\} \to K^0 \pi^+ \pi^-~~.
\eea
$K_{\rm res}^{+}$ and $K_{\rm res}^0$ occur in radiative $B^+$ and $B^0$ decays, 
respectively. No interference is present in the amplitudes for the two final 
states $K^+\pi^0 \pi^0$ and $K^0\pi^0 \pi^0$, which can be produced only 
through $K^{*+}\pi^0$ and $K^{*0}\pi^0$ modes, respectively. We note that, 
in the isospin and narrow resonance width limits, and assuming a 
vanishing $\rho K$ contribution,
the partial decay rates of each of the modes (\ref{chain1}) and (\ref{chain2})
((\ref{chain3}) and (\ref{chain4})) for $K_{\rm res}^{+}$ ($K_{\rm res}^{0}$)
is equal to 4/9 of the total decay branching ratio into $K\pi\pi$.

\section{$B\to K\pi\pi\gamma$ in terms of a photon polarization parameter}

Consider the radiative decays into a given kaon resonance state
$\bar B(b\bar q)\to \bar K^{(i)}_{\rm res}\gamma$.
Let us denote the weak radiative amplitudes 
by $c_L^{(i)}\equiv A(\bar B\to K^{(i)}_{\rm res}\gamma_L)$ and 
$c_R^{(i)}\equiv A(\bar B\to K^{(i)}_{\rm res}\gamma_L)$,
for left and right-polarized photons, respectively. The photon 
polarization in $B\to K^{(i)}_{\rm res}\gamma$ is given by 
\beq
\lambda^{(i)}_\gamma = 
\frac{|c^{(i)}_R|^2-|c^{(i)}_L|^2}{|c^{(i)}_R|^2+|c^{(i)}_L|^2}~~.
\eeq

Let us show that the ratios $|c^{(i)}_R/c^{(i)}_L|$, and therefore 
$\lambda^{(i)}_\gamma$, are equal for all $K$ resonance states, and are 
given by fundamental couplings in the effective weak radiative Hamiltonian. 
The effective Hamiltonian has the general structure
\beq\label{Hrad}
{\cal H}_{\rm rad} = -\frac{4G_F}{\sqrt2} V_{tb} V_{ts}^*\left( C_{7R}{\cal
O}_{7R}
+  C_{7L}{\cal O}_{7L}\right)~,~~~~~{\cal O}_{7L,R} = \frac{e}{16\pi^2} m_b
\bar s\sigma_{\mu\nu}\frac{1 \pm \gamma_5}{2} bF^{\mu\nu}~~,
\eeq
where the Wilson coefficients $C_{7L}$ and $C_{7R}$ describe the amplitudes of 
$b \to s\gamma$ for left and right-handed photons, respectively. Due to the
chiral structure of the $W^\pm$ couplings to quarks in the SM, the amplitude for
the emission of a left-handed photon in $b\to s\gamma$ is enhanced relative to
that for a right-handed photon by $C_{7R}/C_{7L} \simeq  m_s/m_b$. This property
holds even after adding to (\ref{Hrad}) the four-quark operators ${\cal O}_{1,2}$,
which have the same $(V-A)\times (V-A)$ chiral structure as the couplings
producing the dominant penguin operator ${\cal O}_{7L}$. On the other hand,
much larger $C_{7R}/C_{7L}$ ratios are permitted in LR and MSSM extensions of the
SM.
 
Parity invariance of the strong interactions relates the amplitude of $b\to 
s\gamma$ for emitting a left-handed photon through ${\cal O}_{7L}$ to the 
amplitude for emitting a right-handed photon through ${\cal O}_{7R}$,
\beq\label{parity}
\langle K^{(i)R}_{\rm res} \gamma_R|{\cal O}_{7R}|\bar B\rangle =
(-1)^{J_i-1}P_i 
\langle K^{(i)L}_{\rm res} \gamma_L|{\cal O}_{7L}|\bar B\rangle~~,
\eeq
where $J_i$ and $P_i$ are the resonance spin and parity, and 
$K^{(i)R,L}_{\rm res}$ denote states with helicities $\pm 1$.

To prove this relation, let us assume for definiteness that the
photon moves along the $+z$ axis and the $K$ resonance along the
opposite direction. Under a parity transformation ${\cal P}$ the
operators ${\cal O}_{7R}$ and ${\cal O}_{7L}$ transform to each other,
${\cal P} {\cal O}_{7R} {\cal P}^\dagger = {\cal O}_{7L}$. This gives
\bea\label{step1}
\langle K^{(i)R}_{\rm res}(\downarrow) 
\gamma_R(\uparrow) |{\cal O}_{7R}|\bar B\rangle =
\langle K^{(i)R}_{\rm res}(\downarrow) \gamma_R(\uparrow) |
{\cal P}^\dagger {\cal P}{\cal O}_{7R}{\cal P}^\dagger {\cal P}
|\bar B\rangle =
P_i
\langle K^{(i)L}_{\rm res}(\uparrow) 
\gamma_L(\downarrow)|{\cal O}_{7L}|\bar B\rangle~,
\eea
where the arrows denote particle momenta relative to the $z$ direction.
Under a rotation ${\cal R}$, around the $y$ axis through $180^\circ$,
the states transform as ${\cal R}|J,M\rangle = (-1)^{J-M}|J,-M\rangle$.
Applying this rotation to the right-hand side of (\ref{step1}) gives 
Eq.~(\ref{parity}).

Together with the equality 
\beq
\langle K^{(i)L}_{\rm res} \gamma_L|{\cal O}_{7R}|\bar B\rangle =
\langle K^{(i)R}_{\rm res} \gamma_R|{\cal O}_{7L}|\bar B\rangle = 0~~,
\eeq
Eq.~(\ref{parity}) shows that, for a given resonance, the weak amplitudes 
$c^{(i)}_R$ and $c^{(i)}_L$ are proportional, up to a sign, to the Wilson 
coefficients $C_{7R}$ and $C_{7L}$, respectively, and to a common hadronic 
matrix element of ${\cal O}_{7R}$, $g_+^{(i)}(0)$

\bea\label{cRL}
\left\{
\begin{array}{c}
c_R^{(i)} \\
c_L^{(i)}
\end{array}
\right\} = - \frac{4G_F}{\sqrt2} 
\left\{
\begin{array}{c}
C_{7R} \\
P_i (-1)^{J_i-1} C_{7L}
\end{array}
\right\} 
V_{tb} V_{ts}^* g_+^{(i)}(0)~~.
\eea
This implies
\beq\label{lambda}
\frac{|c^{(i)}_R|}{|c^{(i)}_L|} = \frac{|C_{7R}|}{|C_{7L}|} 
~~~\Rightarrow~~~
\lambda^{(i)}_\gamma = \frac{|C_{7R}|^2 - |C_{7L}|^2}{|C_{7R}|^2 + |C_{7L}|^2}
\equiv \lambda_\gamma 
~~.
\eeq

Namely, the photon polarization 
in $\bar B\to \bar K^{(i)}_{\rm res}\gamma$
is common to all $K$ resonance states and 
is given purely in terms of Wilson coefficients.
We will refer to the quantity $\lambda_\gamma$, defined by Wilson
coefficients, as the photon polarization parameter.
We note that the above argument does not depend on the form factor matrix 
elements $g_+^{(i)}(0)$ between $B$ and $K^{(i)}_{\rm res}$, which were 
calculated in several models \cite{gK}.

Now, consider the decays $\bar B(b\bar q)\to \bar K\pi\pi\gamma$, to which 
several overlapping kaon resonances $K^{(i)}_{\rm res}$ contribute. Let us 
denote the strong decay amplitudes for $\bar K^{(i)}_{\rm res}\to \bar K\pi\pi$ 
by $A_L^{(i)}$ and $A_R^{(i)}$, corresponding to a left and 
right-polarized resonance, respectively. The radiative differential decay rate 
can be written as a sum of contributions from left and right-polarized 
photons,
\beq\label{A^2}
d\Gamma(\bar B\to \bar K\pi\pi\gamma) =
\left|\sum_i\frac{c_R^{(i)} A_R^{(i)}}{s-M_i^2-
iM_i\Gamma_i}\right|^2 
+  \left|\sum_i\frac{c_L^{(i)} A_L^{(i)}}{s-M_i^2-
iM_i\Gamma_i}\right|^2~~,
\eeq
where $s = (p_K + p_{\pi_1} + p_{\pi_2})^2$ is the invariant mass of the 
hadronic $K\pi\pi$ state. The two terms do not interfere in the partial decay 
rate since {\it in principle} the photon polarization is measurable. 
Using Eqs.~(\ref{cRL}) and (\ref{lambda}), one finds 
\beq\label{dGamma}
d\Gamma(\bar B\to \bar K\pi\pi\gamma) \propto  
\left( |{\cal A}_R|^2 + |{\cal A}_L|^2 \right ) + 
\lambda_\gamma \left(|{\cal A}_R|^2 - |{\cal A}_L|^2 \right)~~,
\eeq
where
\beq\label{ARL}
{\cal A}_R \equiv \sum_i  g_+^{(i)} A_R^{(i)} B_i(s)~,\qquad
{\cal A}_L \equiv \sum_i P_i (-1)^{J_i-1} g_+^{(i)} A_L^{(i)} B_i(s)~~,
\eeq
and where
\beq\label{BW}
B_i(s) \equiv \frac{1}{s-M_i^2-i\Gamma_i M_i}
\eeq
are Breit-Wigner forms involving the mass $M_i$ and width $\Gamma_i$ of 
a resonance $K^i_{\rm res}$.

The term in the decay distribution (\ref{dGamma}), which is sensitive to 
$\lambda_\gamma$,
involves the diffeence $|{\cal A}_R|^2 - |{\cal A}_L|^2$. For a single
resonance $K^{(i)}_{\rm res}$, it is proportional to $|A_R^{(i)}|^2 -
|A_L^{(i)}|^2$. Therefore, a measurement of the photon polarization parameter
$\lambda_\gamma$ is sensitive to asymmetries between the resonance decay 
amplitudes $|A_R^{(i)}|$ and $|A_L^{(i)}|$. Such a measurement can be directly 
translated into information about the ratio of Wilson coefficients 
$|C_{7R}/C_{7L}|$, which is given in the SM by $m_s/m_b$ implying 
$\lambda_\gamma = -1~(+1) + {\cal O}(m^2_s/m^2_b)$, for $\bar B~(B)$ decays, 
respectively. We will show that such a determination is 
essentially free of hadronic uncertainties, and can be performed even in the 
absence of any information about the strong matrix elements $g_+^{(i)}$.

It is interesting to note the relation between the photon polarization 
parameter $\lambda_\gamma$, which is defined in (\ref{lambda}) in terms of 
Wilson coefficients, and the photon polarization in $\bar B\to \bar 
K\pi\pi\gamma$. The latter can be defined by
\bea
P_\gamma &=& 
\frac{\Gamma(\bar B\to \bar K\pi\pi \gamma_R) - \Gamma(\bar B\to \bar K\pi\pi 
\gamma_L)}
{\Gamma(\bar B\to \bar K\pi\pi \gamma_R) + \Gamma(\bar B\to \bar K\pi\pi 
\gamma_L)}\\
&=& \frac{\int dPS (|{\cal A}_R|^2 - |{\cal A}_L|^2) + \lambda_\gamma
\int dPS (|{\cal A}_R|^2 + |{\cal A}_L|^2)}
{\int dPS (|{\cal A}_R|^2 + |{\cal A}_L|^2) + \lambda_\gamma
\int dPS (|{\cal A}_R|^2 - |{\cal A}_L|^2)}~~,\nonumber
\eea
where $\int dPS$ denotes an integral over entire phase space. 
For a single resonance, one has  
\beq\label{single}
\int dPS (|{\cal A}_R|^2 - |{\cal A}_L|^2) = |g_+^{(i)} B_i(s)|^2 \int dPS
(|A^{(i)}_R|^2 - |A^{(i)}_L|^2)~~,
\eeq
and parity conservation in the resonance decay implies that the right hand side 
vanishes. In this case one would have $P_\gamma = \lambda_\gamma$. 
This is not true for several interfering resonances for which a relation similar 
to (\ref{single}) does not hold in general. Namely, the photon polarization 
parameter $\lambda_\gamma$ coincides with the photon polarization $P_\gamma$ in 
$\bar B \to \bar K\pi\pi\gamma$ only for a process proceeding through a single 
resonance.

\section{Strong decays $K_{\rm res}\to K\pi\pi$}

In the resonance mass region $M_{\rm res} = 1300-1500$ MeV, which we consider,
there exist several $K$ resonances with different quantum numbers $J^P$,
which decay to $K\pi\pi$.
We list these states in Table 1, specifying their masses, widths and
decay branching ratios \cite{PDG}. 

\begin{center}
\begin{tabular}{cccccc}
\hline
Resonance & $J^P$ & $(M_{\rm res},\Gamma_{\rm res})$ (MeV) &
Decay mode & Br (\%)\\
\hline
\hline
$K_1(1270)$ & $1^+$ & $(1273\pm 7\,, 90\pm 20)$ &
$\rho K$ & $42\pm 6$ \\
& & & $K^*\pi$ & $16\pm 5$ \\
& & & $K_0^*(1430)\pi$ & $28\pm 4$ \\
\hline
$K_1(1400)$ & $1^+$ & $(1402\pm 7\,, 174\pm 13)$ & $K^*\pi$ & $94\pm 6$ \\
& & & $\rho K$ & $3.0\pm 3.0$ \\
\hline
$K^*(1410)$ & $1^-$ & $(1414\pm 15\,, 232\pm 21)$ & $K^*\pi$ & $>40$ \\
& & & $\rho K$ & $<7$ \\
\hline
$K_2^*(1430)$ & $2^+$ & $(1425.6\pm 1.5 \,, 98.5\pm 2.7)$ & $K^*\pi$ & 
$24.7\pm 1.5$ \\
& & (charged $K_2^*$) & $\rho K$ & $8.7\pm 0.8$ \\
\hline
\hline
\end{tabular}
\end{center}
\begin{quote} {\bf Table I.}
Kaon resonances with masses in the region 1250 - 1450 MeV decaying into 
$K\pi\pi$. 
\end{quote}

The strong decay of a kaon resonance $K_{\rm res}$ into a 3-body $K\pi\pi$ 
final state proceeds through the graphs shown in Fig.~1, with intermediate 
$K^*\pi$ and $\rho K$ states. In the following subsections we use these diagrams 
to compute for the three resonances around 1400 MeV the strong decay 
amplitudes $A^{(i)}_L$ and $A^{(i)}_R$ appearing in the rate equations  
(\ref{dGamma}) and (\ref{ARL}).

\subsection{Decays of $K_1(1400)$ ($J^P = 1^+$)}

The $K_1(1400)$ decays predominantly to $K^*(892)\pi$ with a branching
ratio of $(94\pm 6)\%$, and to $\rho K$ with a much smaller branching ratio,
$(3.0\pm 3.0)\%$. Both decays occur in a mixture of $S$
and $D$ waves. The $D$ to $S$ wave ratio of rates in the $K_1\to K^*(892)\pi$ 
channel was measured to be small, $0.04\pm 0.01$ \cite{PDG,Daum}. No similar 
measurement exists for the $\rho K$ channel.

\begin{figure}[t]
\centerline{\epsfysize = 1.2 in \epsffile{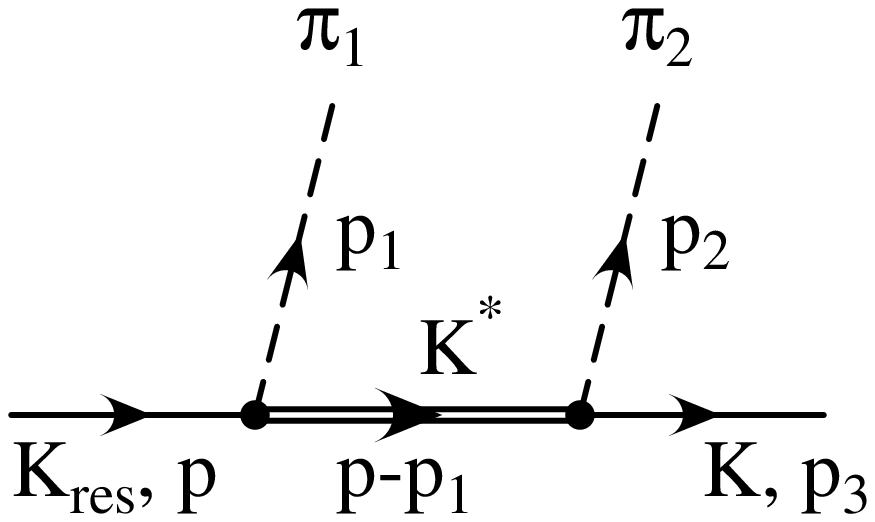}
\epsfysize = 1.2 in \epsffile{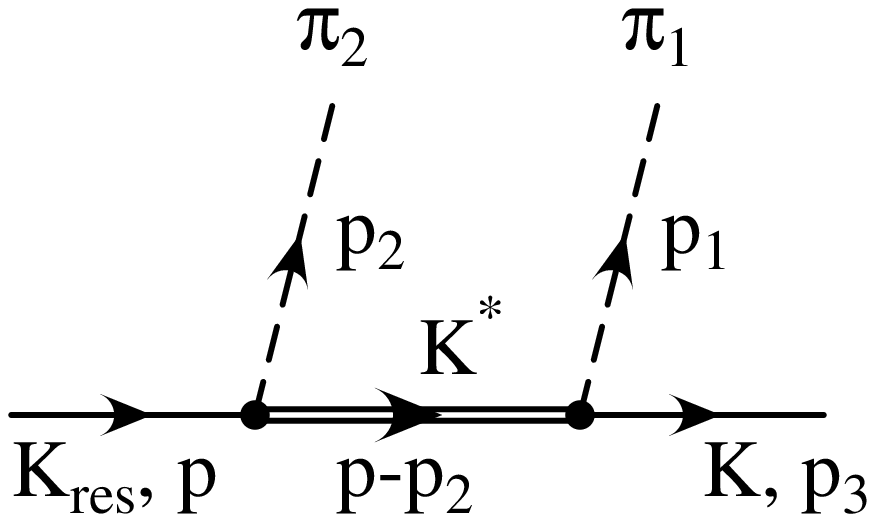}
\epsfysize = 1.2 in \epsffile{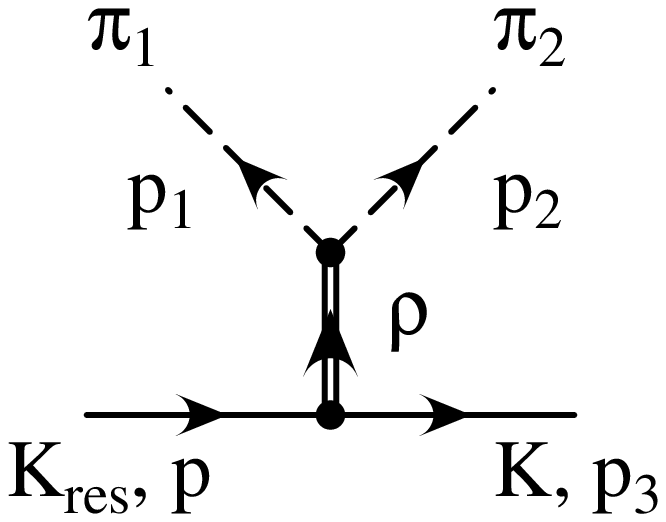}}
\centerline{(a)\hspace{4.5cm} (b)\hspace{4.5cm} (c)}
\caption{Resonant contributions to the decay $K_{\rm res}\to K\pi\pi$,
proceeding through $K^*\pi$ and $\rho K$ intermediate states.}
\end{figure}

The invariant matrix element for $K_1 \to K^*\pi$ (and similarly for 
$K_1 \to \rho K$) can be parametrized in terms of two complex couplings,
$A^{(K^*)}$ and $B^{(K^*)}$,
\beq\label{AB}
{\cal A}(K_1(p,\varepsilon)\to K^*(p_1, \varepsilon') \pi(p_\pi)) =
A^{(K^*)} (\varepsilon\cdot \varepsilon^{'*}) + B^{(K^*)} (\varepsilon^{'*}\cdot p_\pi)
(\varepsilon\cdot p_\pi)~~,
\eeq
where momenta and polarization 4-vectors are specified for each of the 
particles. The couplings $A^{(K^*)}$ and $B^{(K^*)}$ are related to the 
partial wave amplitudes $c_S$ and $c_D$ through
\bea\label{cSD}
A^{(K^*)} &=& c_S + c_D \frac{M_{K_1} \vec p_\pi\,^2}{2M_{K^*}+E_{K^*}}~,\qquad
B^{(K^*)} = c_D + c_S \frac{E_{K^*}-M_{K^*}}{M_{K_1} \vec p_\pi\,^2}~~.
\eea
In the nonrelativistic limit, $M_{K_1}-M_{K^*} \ll M_{K_1}$, this reduces
to the more familiar couplings, $c_S(\vec \varepsilon\cdot \vec 
\varepsilon\,^{'*}) + c_D [(\vec \varepsilon\cdot \vec p_\pi) (\vec 
\varepsilon\,^{'*}\cdot \vec p_\pi) - \frac13 \vec p_\pi\,^2 (\vec 
\varepsilon\cdot \vec \varepsilon\,^{'*})]$.
[A similar partial wave decomposition was performed in \cite{Feindt} within 
the context of $a_1\to\rho\pi$ decays.]
Leaving out isospin factors, the amplitude (\ref{AB}) implies the following
result for the  $\Gamma(K_1\to K^*\pi)$ width 
\bea\label{K1rate}
\Gamma(K_1\to K^*\pi) = |c_S|^2 \frac{|\vec p_\pi\,|}{8\pi M_{K_1}^2} +
|c_D|^2 \frac{|\vec p_\pi\,|^5}{4\pi (E_{K^*} + 2M_{K^*})^2}~~,
\eea
where the two terms correspond to the $S$-wave and $D$-wave partial widths, 
respectively.
A similar parametrization can be given for the $K_1\to K\rho$ coupling.

Using the rate formula (\ref{K1rate}) and the measured value of the 
$D$ to $S$ ratio of width, one finds $|c_D/c_S| = (1.75\pm 0.22)$ GeV$^{-2}$. 
The relative phase between the $D$ and $S$ wave amplitudes was measured 
in  \cite{Daum} to be $\delta_{D/S} \equiv {\rm Arg}\,(c_D/c_S) = 260^\circ 
\pm 20^\circ$. These values can be used to determine $A^{(K^*)}$ and 
$B^{(K^*)}$ from Eq.~(\ref{cSD}). 

The amplitude for $K_1(1400)\to K\pi\pi$ is obtained from the graphs shown 
in Fig.~1 by convoluting amplitudes such as (\ref{AB}) with the 
amplitude for $K^*\to K\pi$ which is proportional to $\varepsilon'\cdot 
(p_\pi - p_K)$. The modes (\ref{chain1})~and~(\ref{chain3}) obtain 
contributions from all three graphs Figs.~1(a)-(c), while in the 
modes (\ref{chain2}) and (\ref{chain4}) only the graphs Fig.~1(a) 
and Fig.~1(c) contribute. In the first case, 
isospin considerations imply that the two $K^*$ contributions from
Fig.~1(a) and Fig.~1(b) are antisymmetric under 
the exchange of the two pion momenta. We will write down general expressions for
(\ref{chain1}) and (\ref{chain2}), noting that the amplitudes of the processes 
(\ref{chain3}) and (\ref{chain4}) have correspondingly similar forms.

The amplitude for the process (\ref{chain1}) can be summarized in the rest 
frame of the $K_1$ by the following expression
\beq\label{M1plus}
{\cal M}(K_1^+(p,\varepsilon)\to \pi^+(p_1)\pi^0(p_2) K^0(p_3)) =
C_1 \vec p_1\cdot \vec \varepsilon - C_2 \vec p_2\cdot \vec \varepsilon~~,
\eeq
where 
\beq\label{Ci}
C_i(s_{13}, s_{23}) = C^{(K^*)}_i(s_{13}, s_{23}) + C^{(\rho)}_i(s_{13}, s_{23})
~,~~~s_{ij} = (p_i + p_j)^2~~.
\eeq
The explicit expressions for $C^{(K^*)}_i$ and $C^{(\rho)}_i$ are:
\bea\label{CK*rho}
& & C_1^{(K^*)}(s_{13}, s_{23}) = \frac{\sqrt2}{3}g_{K^*K\pi} 
A^{(K^*)} \left[ \left(1+\frac{m_2^2-m_3^2}{M_{K^*}^2}\right) B_{K^*}(s_{23}) - 
2B_{K^*}(s_{13}) \right] \nonumber \\
 &+& \frac{\sqrt2}{3} g_{K^*K\pi} B^{(K^*)} \left[
-\left(1+\frac{m_2^2-m_3^2}{M_{K^*}^2}\right)(M_{K_1} E_1 - m_1^2) + 
2p_1\cdot p_2 \right] B_{K^*}(s_{23})~~,\nonumber \\
& & C_2^{(K^*)}(s_{13}, s_{23}) = \frac{\sqrt2}{3} g_{K^*K\pi} 
A^{(K^*)} \left[ \left(1+\frac{m_1^2-m_3^2}{M_{K^*}^2}\right) B_{K^*}(s_{13}) - 
2B_{K^*}(s_{23}) \right]\nonumber \\
 &+& \frac{\sqrt2}{3} g_{K^*K\pi} B^{(K^*)} \left[
-\left(1+\frac{m_1^2-m_3^2}{M_{K^*}^2}\right)(M_{K_1} E_2 - m_2^2) + 
2p_1\cdot p_2 \right] B_{K^*}(s_{13})~~,\nonumber \\
& & C_1^{(\rho)}(s_{13}, s_{23}) = \frac{1}{\sqrt3} g_{\rho\pi\pi} A^{(\rho)}  
B_\rho(s_{12}) - \frac{1}{\sqrt3} g_{\rho\pi\pi} B^{(\rho)} M_{K_1} (E_1-E_2) 
B_\rho(s_{12})~~,\nonumber \\
& & C_2^{(\rho)}(s_{13}, s_{23}) = \frac{1}{\sqrt3} g_{\rho\pi\pi} A^{(\rho)}  
B_\rho(s_{12}) + \frac{1}{\sqrt3} g_{\rho\pi\pi} B^{(\rho)} M_{K_1} (E_1-E_2) 
B_\rho(s_{12})~~,\nonumber \\
& & \,
\eea
where $B_{K^*}(s_{ij})$ and $B_{\rho}(s_{ij})$ are Breit-Wigner functions
\beq\label{BW2}
B_{K^*}(s_{ij}) = \left( s_{ij} - M_{K^*}^2 - iM_{K^*} \Gamma_{K^*}
\right)^{-1}~,~~~
B_{\rho}(s_{ij}) = \left( s_{ij} - m_{\rho}^2 - im_{\rho} \Gamma_{\rho}
\right)^{-1}~.
\eeq

The corresponding amplitudes for the modes (\ref{chain2}) and (\ref{chain4}) 
are obtained from (\ref{M1plus}) by setting $B_{K^*}(s_{13})\to 0$,
and multiplying the $K^*\pi$ part by $-\sqrt2$ and the $K\rho$ part by 
$-1/\sqrt2$. All amplitudes are symmetric under the exchange of the two pions,
as expected from Bose symmetry.

Whereas the $K_1(1400)$ decays predominantly to $K^*\pi$, the small measured
branching ratio into $\rho K$,~$3 \pm 3 \%$, which only implies an upper limit,
may have a 
nonnegligible effect on the decay distribution from which the photon
polarization is measured. In order to estimate this effect, we parametrize
the relative contributions of the $\rho K$ and $K^*\pi$ amplitudes in terms 
of two complex ratios of corresponding partial wave amplitudes for $S$ and $D$ 
waves, 
\beq\label{kapparhoK1}
\kappa_{S,D} = |\kappa_{S,D}| e^{i\alpha_{S,D}} =
\sqrt{\frac32} \frac{c_{S,D}^{(\rho K)}}{c_{S,D}^{(K^*\pi)}}\cdot
\frac{g_{\rho\pi\pi}}{g_{K^* K\pi}}~~.
\eeq
In the absence of experimental data on the individual partial widths for 
the $\rho K$ mode, these parameters cannot be determined at present.
Some measure for the $\rho K$ contribution can be 
obtained by neglecting the $D$-wave admixture in this mode and assuming that 
the measured width (for which only an upper limit exists) is pure $S$-wave. 
With this approximation, the coefficients $C_i$ in (\ref{M1plus}) are 
given by
\bea\label{C1}
C_1(s_{13}, s_{23}) &\propto& (1+0.081 \frac{c_D}{c_S})
\left[ \left(1+\frac{m_2^2-m_3^2}{M_{K^*}^2}\right) B(s_{23}) - 
2B(s_{13})\right]\\
&+& (0.384 +\frac{c_D}{c_S})
 \left[
-\left(1+\frac{m_2^2-m_3^2}{M_{K^*}^2}\right)(M_{K_1} E_1 - m_1^2) + 
2p_1\cdot p_2
\right] B(s_{23}) \nonumber\\
&+& \kappa_S [1 - 0.631 (E_1-E_2)] B_\rho(s_{12})\nonumber~~,
\eea
and $C_2(p_1,p_2) = C_1(p_2,p_1)$.

The absolute value of the ratio $\kappa_S$ can be obtained from the 
measured widths for the respective modes
\bea\label{ratio1}
\frac{|c_{S}^{(\rho K)}|^2}{|c_S^{(K^*\pi)}|^2}
 &=& \frac{{\cal B}_{S}(K_1\to \rho K)}{{\cal B}_{S}(K_1\to K^*\pi)}\cdot
\frac{|\vec p_{K^*\pi}\,|}{|\vec p_{\rho K}\,|}\simeq 0.043~~,\\
\label{ratio2}
\frac{|g_{\rho\pi\pi}|^2}{|g_{K^* K\pi}|^2}
 &=& \frac{2\Gamma_\rho}{\Gamma_{K^*}}\cdot
\frac{|\vec p_{K\pi}\,|^3}{|\vec p_{\pi\pi}\,|^3} = 3.16~~,
\eea
where we assumed that the measured $\rho K$ branching ratio is pure $S$-wave
and is given by the central value. Thus we find from Eq.~(\ref{kapparhoK1}) 
$|\kappa_S| = 0.45$. 
The relative phase $\alpha_S$ was measured in \cite{Daum}, and was found to
lie in the range $20^\circ \le \alpha_S \le 60^\circ$.
In our subsequent numerical calculation we will use the range $|\kappa_S| = 
0.32 \pm 0.32$, corresponding to ${\cal B}(K_1 \to \rho K) = 0.03 \pm 0.03$, 
and will scan the values of the phase $\alpha_S$ in the above range.

\subsection{Decays of $K_2^*(1430)$ ($J^P = 2^+$)}

The $K_2^*(1430)$ decays to $K^*\pi$ and $\rho K$ in 
a pure $D$-wave, with branching ratios of $(24.7\pm 1.5)\%$ and $(8.7\pm 0.8)\%$ 
respectively. The invariant amplitude for a $K^*\pi$ final state is written in 
terms of a single coupling,
\beq
{\cal A}(K_2^*(p,\varepsilon) \to K^*(p_1,\varepsilon_1)\pi(p_2)) =
g_{K_2^* K^*\pi} i\epsilon_{\alpha\beta\gamma\delta} \varepsilon_1^{*\alpha}
p^\beta \varepsilon^{\gamma\rho} p_{2\rho} p_2^\delta~~,
\eeq
and a similar expression can be written for ${\cal A}(K_2^* \to \rho K)$.
The amplitude for $K_2^*\to K\pi\pi$ obtains contributions from the three
graphs in Fig.~1. Their computation gives for charged $K^*_2$ 
decays of the type (\ref{chain1}):
\bea\label{M2plus}
& &{\cal M}(K_2^{*+}(p,\varepsilon)\to \pi^+(p_1)\pi^0(p_2)K^0(p_3)) \\
& &\quad \propto i\epsilon_{\alpha\beta\gamma\mu}
p^\alpha p_1^\beta p_2^\gamma \varepsilon^{\mu\nu}
\left[p_{1\nu} B_{K^*}(s_{23})  + p_{2\nu} B_{K^*}(s_{13}) +
\kappa(p_{1\nu} + p_{2\nu})  B_\rho(s_{12}) \right]~~.\nonumber
\eea
A similar expression is obtained for neutral $K^*_2$ decays of the type
(\ref{chain3}). Here again, Bose symmetry requires the amplitude to be 
symmetric under an exchange of the two pions.

The ratio of the $\rho K$ and $K^*\pi$ contributions is parametrized by a
complex parameter $\kappa$, which is defined in a way similar to 
Eq.~(\ref{kapparhoK1}),
\beq\label{kapparhoK2}
\kappa = |\kappa| e^{i\alpha} =
\sqrt{\frac32} \frac{g_{K_2^*\rho K}}{g_{K_2^*K^*\pi}}\cdot
\frac{g_{\rho\pi\pi}}{g_{K^* K\pi}}~~.
\eeq
Using the measured branching ratios of these modes, one finds for 
the absolute value of the first ratio,
\bea
\frac{|g_{K_2^*\rho K}|^2}{|g_{K_2^*K^*\pi}|^2}  =
\frac{{\cal B}(K_2^*\to \rho K)}{{\cal B}(K_2^*\to K^* \pi)}\times
\frac{|\vec p_{K^*\pi}\,|^5}{|\vec p_{\rho K}\,|^5} = 1.2~~,
\eea
which gives $|\kappa| = 2.38$ when applying Eq.~(\ref{ratio2}). 

We will argue now that also the phase $\alpha$ of $\kappa$ can be constrained 
using available experimental data. Let us consider first the 
phase $\alpha_1$ of $g_{\rho\pi\pi}/g_{K^* K\pi}$, the second factor in 
$\kappa$. This ratio, given in Eq.~(\ref{ratio2}), is predicted to be 
$\sqrt{8/3}$ in the SU(3) limit. The small SU(3) breaking in the measured value 
of this ratio (about 8\%) suggests that its phase is also small. 
Now consider the first ratio in $\kappa$, $g_{K_2^*\rho K}/g_{K_2^*K^*\pi}$. 
Although SU(3) symmetry does not predict this ratio, it is possible to 
determine its phase $\alpha_2$ by noting that in the SU(3) limit the amplitudes 
for the decays $K_2^*\to K^* \pi$, $K_2^*\to \rho K$ and $a_2(1320)\to 
\rho \pi$ satisfy a triangle relation,
\beq\label{triangleK2}
A(K_2^{*+}\to K^{*+} \pi^0) + \frac{1}{\sqrt2} A(K_2^{*+}\to \rho^{+} K^0) =
A(a_2^{+}(1320)\to \rho^{+} \pi^0)~~.
\eeq
$\alpha_2$ is given by the relative phase of the amplitudes on the
left-hand side. Using the measured widths for these modes one finds that,
although the triangle does not close at the central values of the measured 
amplitudes (the right-hand side of Eq.~(\ref{triangleK2}) is slightly larger 
than the algebraic sum of the amplitudes on the left hand-side), it closes 
with a very small angle $\alpha_2$ when errors are included.
Namely, $\alpha_2$ is close to zero in the SU(3) limit. Allowing for some
SU(3) breaking effects, we will use in our numerical estimates below values for
$\alpha = \alpha_1 + \alpha_2$ between $-30^\circ$ and $+30^\circ$. A small
phase in this range was measured in a $K^*_2$ resonance production experiment 
\cite{Daum}.

\subsection{Decays of $K^*(1410)$ $(J^P=1^-)$}

The $K_1^*(1410)$ decays predominantly to $K^*\pi$ in a
pure $P$-wave, with a branching ratio larger than 40\% (at 95\% CL), while an 
upper bound of 7$\%$ exists for its decay branching ratio into $\rho K$.
The invariant amplitude describing the first decay is
\beq
{\cal A}(K_1^*(p,\varepsilon)\to K^*(p_1,\varepsilon_1) \pi(p_2)) = 
g_{K_1^* K^*\pi} i\epsilon(p,\varepsilon,\varepsilon_1^*,p_2)~~.
\eeq
Using Figs.~1 to calculate three contributions, one finds  
the $K_1^*\to K\pi\pi$ amplitude in the rest frame of the decaying resonance,
\bea\label{K*1}
{\cal M}(K_1^*(p,\varepsilon)\to \pi(p_1)\pi(p_2)K) \propto 
\vec \varepsilon\cdot (\vec p_1\times \vec p_2) (B_{K^*}(s_{13}) +
B_{K^*}(s_{23}) + \kappa' B_\rho (s_{12}))~,
\eea
where $\kappa'$ parametrizes the ratio of $\rho K$ and $K^*\pi$ contributions.
The information on the magnitude of this ratio from measured branching ratios 
is limited to a rather weak upper bound. 

It is quite simple to argue from a general principle that the decay amplitude 
(\ref{K*1}) leads to a radiative decay distribution which is insensitive to the 
photon polarization. The only parity invariant decay amplitude
which can be constructed from the $K^*_1$ polarization vector 
$\vec \varepsilon$ and the final mesons momenta, is proportional
to $\vec \varepsilon\cdot (\vec p_1\times \vec p_2)$. Its square is invariant
under $\vec \varepsilon_{+1}\leftrightarrow \vec \varepsilon_{-1}$ and 
therefore cannot be used to measure the photon polarization. 

\section{Angular distributions}

As explained in Sec.~III, Eqs.~(\ref{dGamma}) -- (\ref{ARL}) and the subsequent
discussion, the sensitivity to the photon polarization parameter is
manifested through an asymmetry between the decay distributions of right and 
left-polarized $K$ resonance states.
In order to compute this asymmetry, one has to specify the orientation of the 
decay products $K\pi\pi$ with respect to the so-called helicity axis, defined
to be along the photon momentum direction and opposite to it 
$\hat e_z = -\vec p_\gamma$. Working in the rest frame of the $K\pi\pi$ state,
we define the normal $\vec n$ to 
the $K\pi\pi$ plane as $\vec n = (\vec p_1\times \vec p_2)/
|\vec p_1\times \vec p_2\,|$. The orientation of the $K\pi\pi$ system with 
respect to the helicity axis $\vec e_z$ is given in the most general case by
three angles $(\theta,\phi,\psi)$.
Two polar angles $(\theta,\phi)$, with $\cos\theta=\vec n \cdot \vec e_z$, 
describe the orientation of $\vec n$ with
respect to $\vec e_z$. A third angle $\psi$ (unobservable in this case)
parametrizes rotations around the $\vec e_z$ axis. In the subsequent study 
we will derive angular distributions in the angle $\theta$, integrating over 
the azimuthal angle $\phi$.

In Ref.~\cite{ggpr} angular distributions were studied for $B \to K\pi\pi
\gamma$ separately for the
three resonances $K_1(1400),~K^*_2(1430)$ and $K^*(1410)$.
Here we will study the most general decay distribution combining all three 
overlapping resonances including their interference.
The structure of the amplitude of $\bar B \to \bar K\pi\pi\gamma_{R,L}$,
given in Eq.~(\ref{ARL}) for right
and left-polarized photons, can be obtained by summing over 
contributions calculated in the preceding section for the three $K$ resonance 
states. One finds, in the rest frame of the $K\pi\pi$ system,
\beq\label{A-1}
{\cal A}_{R,L}(\bar B\to \bar K\pi\pi\gamma_{R,L}) =  A(\vec 
\varepsilon_\pm\cdot 
\vec J) \pm B\left( (\vec \varepsilon_\pm\cdot \vec n )(\vec 
\varepsilon_0\cdot \vec K) 
+ (\vec \varepsilon_\pm\cdot \vec K )(\vec \varepsilon_0\cdot \vec n)\right)
\pm C(\vec \varepsilon_\pm\cdot \vec n)~,
\eeq
where the polarization vectors $\vec \varepsilon_i$ are defined in 
terms of $\vec e_z$ and two arbitrary unit vectors $\vec e_x$ and $\vec e_y$ 
in the plane perpendicular to $\vec e_z$,
\beq\label{unitvectors}
\vec \varepsilon_\pm = \mp\frac{1}{\sqrt2}(\vec e_x \pm i \vec e_y)~,\qquad
\vec \varepsilon_0 = \vec e_z~~.
\eeq

The three terms in (\ref{A-1}) are obtained from intermediate $K_{\rm res}$ 
states with quantum numbers $J^P = 1^+, 2^+$ and $1^-$, respectively.
Their strong decay amplitudes were given in Eqs.~(\ref{M1plus}), (\ref{M2plus}) 
and  (\ref{K*1}), respectively. In order to obtain the second term from 
Eq.~(\ref{M2plus}), we used the following expressions for polarization tensors 
corresponding to right and left-handed $K_2^*$ of helicity $\pm 1$:
\bea\label{eps}
\varepsilon_{\pm 1}^{\mu\nu} = \frac{1}{\sqrt2} (\varepsilon_{\pm 1}^\mu
\varepsilon_{0}^\nu + \varepsilon_{0}^\mu \varepsilon_{\pm 1}^\nu)~~,
\eea
where $\varepsilon^0_m=0$ ($m = \pm1, 0$).

The coefficients $A,B,C$ include the strong matrix elements $g_+^{(i)}$ 
and the Breit-Wigner forms defined in (\ref{BW}).
The vectors $\vec J, \vec K$ (lying in the decay 
plane of $K\pi\pi)$ are functions of the Dalitz variables $s_{13}, s_{23}$. 
Explicit expresions for $\vec J$ and $\vec K$ are obtained from 
Eqs.~(\ref{M1plus}) and (\ref{M2plus}), respectively,
\bea\label{JK}
\vec J &=& C_1 \vec p_1 - C_2 \vec p_2~~,\\
\vec K &=& |\vec p_1\times\vec p_2\,|\left\{\vec p_1 [B_{K^*}(s_{23}) + 
\kappa_\rho B_\rho(s_{12})] +
\vec p_2 [B_{K^*}(s_{13}) + \kappa_\rho B_\rho(s_{12})]\right\}~~.\nonumber
\eea
In the limit of isospin symmetry $\vec K$ is symmetric under 
$s_{13} \leftrightarrow s_{23}$, while $\vec J$ changes sign. 

The dot products in (\ref{A-1}) can be expressed in terms of the 
angles $(\theta,\phi,\psi)$ described above, 
\bea
\vec \varepsilon_\pm\cdot \vec n &=& -\frac{i}{\sqrt2} \sin\theta 
e^{\mp i\psi}~,
\qquad
\vec \varepsilon_0 \cdot \vec n = \cos\theta~~, \nonumber\\
\vec \varepsilon_\pm\cdot \vec J &=&
\mp\frac{1}{\sqrt2} e^{\mp i\psi}\left[
(\cos\phi J_x + \sin\phi J_y) \mp i\cos\theta
(\sin\phi J_x - \cos\phi J_y)\right]~~,\nonumber\\
\vec \varepsilon_0 \cdot \vec J &=&
\sin\theta (\sin\phi J_x - \cos\phi J_y)~~.
\eea
Similar expressions hold for $\vec \varepsilon_\pm\cdot \vec K$ and 
$\vec \varepsilon_0 \cdot \vec K$.

Squaring the amplitude (\ref{A-1}), and integrating over $\phi$, one finds
\bea\label{A-2}
& & \frac{1}{2\pi}\int d\phi
|{\cal A}_{R,L}|^2 = |A|^2 \left\{\frac14 |\vec J\,|^2 (1+\cos^2\theta) \pm
\frac12 \cos\theta \mbox{Im}[\vec n\cdot (\vec J\times\vec J\,^*)]\right\}
\nonumber\\
&+& |B|^2 \left\{
\frac14 |\vec K\,|^2 (\cos^2\theta + \cos^2 2\theta ) \pm 
\frac12 \cos\theta\cos 2\theta \mbox{Im}[\vec n\cdot (\vec K\times\vec K\,^*)]
\right\} + |C|^2 \frac12 \sin^2\theta\nonumber \\
&+& \left\{\frac12 (3\cos^2\theta-1) 
\mbox{Im}[AB^* \vec n\cdot (\vec J\times \vec K\,^*)]
\pm \cos^3\theta \mbox{Re}[AB^*(\vec J\cdot \vec K^*)] 
\right\}~~.
\eea
The interference terms between the $1^-$ and the other resonances vanish 
identically upon integration over $\phi$, but there remains a nonvanishing 
interference
between the $J^P=1^+$ and $2^+$ contributions, manifested in the last term.

The decay distribution for $\bar B\to \bar K\pi\pi\gamma$ is readily obtained 
as a function of the photon polarization parameter $\lambda_\gamma$,
\bea\label{dist}
& & \frac{d\Gamma}{ds_{13}ds_{23}d\cos\theta} =
|A|^2 \left\{\frac14 |\vec J\,|^2 (1+\cos^2\theta) +
\frac12 \lambda_\gamma \mbox{Im}[\vec n\cdot (\vec J\times\vec J\,^*)]
\cos\theta \right\}\\
&+& |B|^2 \left\{
\frac14 |\vec K\,|^2 (\cos^2\theta + \cos^2 2\theta ) + 
\frac12 \lambda_\gamma \mbox{Im}[\vec n\cdot (\vec K\times\vec K\,^*)]
\cos\theta\cos 2\theta\right\} + |C|^2 \frac12 \sin^2\theta\nonumber \\
&+& \left\{\frac12 (3\cos^2\theta-1) \mbox{Im}[AB^* \vec n\cdot (\vec 
J\times \vec K\,^*)]
+ \lambda_\gamma \mbox{Re}[AB^*(\vec J\cdot \vec K^*)] \cos^3\theta 
\right\}~~.\nonumber
\eea
This decay distribution is sensitive to $\lambda_\gamma$ 
through the second terms in the $|A|^2$, $|B|^2$ and in the interference 
contributions. Each of these three terms introduces an asymmetry between the 
decay rates for right and left-polarized photons. 
This asymmetry, from which $\lambda_\gamma$ can be determined, describes
an up-down asymmetry of the photon momentum with respect to the $K\pi\pi$ decay 
plane.

Noting the symmetry properties of $\vec J$ (odd) and $\vec K$ (even) under 
the exchange
of $s_{13}$ and $s_{23}$, one can see that after integrating over the entire 
Dalitz plot (or any symmetric part of it), the asymmetry terms proportional to 
$|A|^2$, $|B|^2$ average to zero. In order to avoid this loss of 
polarization information, we introduce a new 
angle $\tilde\theta$ defined by $\cos\theta \equiv 
\mbox{sgn}(s_{13}-s_{23})\,\cos\tilde\theta$. An equivalent definition of 
$\tilde\theta$ is the angle between $-\vec p_\gamma$ and the normal
to the decay plane defined by $\vec p_{\rm slow}\times \vec p_{\rm fast}$,
where  $\vec p_{\rm slow}$ and $\vec p_{\rm fast}$ are the momenta of the
slower and faster pions in the $K\pi\pi$ center of mass frame.

Expressed in terms of $\tilde\theta$, the radiative decay distribution 
(\ref{dGamma}), integrated over a region of the Dalitz plot, has the following 
general form
\bea\label{3res}
& & \frac{\mbox{d}^2\Gamma}{\mbox{d}s \mbox{d}\cos\tilde\theta} = \frac14 
|c_1|^2 |B_{K_1}(s)|^2 \langle |\vec J\,|^2\rangle \left\{
1 + \cos^2\tilde \theta + 4 \lambda_\gamma R_1 \cos\tilde \theta \right\}\\
&+& \frac14 |c_2|^2 |B_{K^*_2}(s)|^2 \langle |\vec K\,|^2\rangle \left\{
\cos^2\tilde\theta + \cos^2 2\tilde\theta + 
12 \lambda_\gamma R_2 \cos\tilde\theta \cos 2\tilde\theta\right\}
 + |c_3|^2 B_{K^*_1}(s) \sin^2\tilde\theta\nonumber\\
&+& 
\left\{\mbox{Im }[c_1 c_2^* B_{K_1}(s) B^*_{K^*_2}(s) c'_{12}]
 \frac12(3\cos^2\tilde\theta -1) + 
\lambda_\gamma \mbox{Re }[c_1 c_2^* B_{K_1}(s) B^*_{K^*_2}(s) c_{12}]
 \cos^3\tilde\theta \right\}\nonumber\,,
\eea
where $\langle\cdots \rangle$ denotes integration over a region of the Dalitz 
plot.
After integrating over $\tilde\theta$, the interference terms in 
the last line vanish, and the rate is given simply by a sum of Breit-Wigner 
forms corresponding to the three $K$ resonance states,
\bea\label{3res1dim}
\frac{\mbox{d}\Gamma}{\mbox{d}s} &=& \frac23 |c_1|^2 \langle |\vec J\,|^2\rangle
|B_{K_1}(s)|^2 + \frac25 |c_2|^2 \langle |\vec K\,|^2\rangle
|B_{K^*_2}(s)|^2 + \frac43 |c_3|^2 |B_{K^*_1}(s)|^2~~.
\eea

In (\ref{3res}) we introduced explicit expressions for $A,B,C$ in terms of 
Breit-Wigner forms with the mass and width of the $K$ resonances with quantum 
numbers $J^P = 1^+, 2^+$ and $1^-$, respectively
\bea
A(s) = c_1 B_{K_1}(s)~,\quad 
B(s) = c_2 B_{K^*_2}(s)~,\quad 
C(s) = c_3 B_{K^*_1}(s)~~,
\eea
where $c_i \propto \langle K^{(i)L}_{\rm res} 
\gamma_L|{\cal O}_{7L}|\bar B\rangle$ are proportional to the decay amplitude 
into the resonance states $K^{(i)}_{\rm res}$.

The coefficients $R_1$ and $R_2$ multiplying the polarization-sensitive terms 
in the decay distributions are defined 
such that $\langle \cos\tilde \theta\rangle = R_1\lambda_\gamma$ and
$\langle \cos\tilde \theta\rangle = R_2\lambda_\gamma$ for decay 
rates dominated by the $J^P = 1^+$ or the $2^+$ resonances, respectively. 
$R_1$ and $R_2$ are expressed in terms of Dalitz plot averages,
\bea\label{Ri} 
R_1 &=& \frac12\cdot\frac{\langle \mbox{sgn}(s_{13}-s_{23})
{\rm Im}\left (\hat n\cdot (\vec J\times \vec J^*)\right )\rangle}
{\langle |\vec J\,|^2\rangle }~~,\\
R_2 &=& \frac16\cdot\frac{\langle \mbox{sgn}(s_{13}-s_{23})
{\rm Im}\left (\hat n\cdot (\vec K\times \vec K^*)\right )\rangle}
{\langle |\vec K\,|^2\rangle }~~,\nonumber
\eea
and are calculable quantities (see discussion below).
The complex coefficients $c_{12}(s)$ and $c'_{12}(s)$, appearing in the 
expression for the $1^+-2^+$ interference terms, are given by
\bea
c_{12}(s) = \langle \mbox{sgn }(s_{13}-s_{23}) \vec 
J\cdot \vec K\,^*\rangle~,\qquad
c'_{12}(s) = \langle \vec n\cdot (\vec J\times \vec K\,^*)\rangle~~.
\eea

Now let us describe a possible procedure which allows a measurement of the 
photon polarization parameter $\lambda_\gamma$ 
in the presence of the three interfering $K$ resonance contributions.
First, one has to determine the three coefficients $|c_i|~ (i=1,2,3)$ 
parametrizing the relative radiative decay rates into the three resonance 
states. In principle, this can be achieved by fitting data to the $K\pi\pi$ 
invariant mass distribution (\ref{3res1dim}), consisting of the sum of three 
Breit-Wigner forms. However, this may be difficult in practice, since the 
masses of the three resonances are too close to each other, and statistics may 
be insufficient for decomposing the destribution into a sum of three different 
Breit-Wigner widths. A complementary way of determining the coefficients $|c_i|$ 
for the three resonances is then to use their spin-parity characteristic angular decay 
distributions into two body or quasi two body states \cite{CLEO, Ishikawa}. 
This analysis can be applied to $K\pi$ final states for measuring $|c_2|$ and 
$|c_3|$ and to $K^*\pi$ states for measuring $|c_1|,~|c_2|$ and $|c_3|$.

Next, we write the angular distribution (\ref{3res}), integrated over a range 
in $s$ centered around the $K_1(1400)$ resonance, as a sum over the seven
functions of $\tilde\theta$ appearing in (\ref{3res}),
\bea\label{ftheta}
F(\tilde \theta) \equiv \int_{s_{\rm min}}^{s_{\rm max}} \mbox{d}s 
\frac{\mbox{d}^2\Gamma}{\mbox{d}s \mbox{d}\cos\tilde\theta} =
\sum_{i=1}^7 d_i f_i(\tilde \theta)~~,
\eea
where $f_1(x) = 1+\cos^2 x\,, f_2(x) = \cos x\,, f_3(x) = \cos^2 x+\cos^2 2x\,,
f_4(x) = \cos 2x \cos x\,,$ $f_5(x) = \sin^2 x\,, f_6(x) = 
\frac12 (3\cos^2 x-1)\,, f_7(x) = \cos^3 x$. The three coefficients 
$d_{1,3,5}$ are obtained from
the corresponding values of $|c_{1,2,3}|$. For example,
\bea
d_1 = \frac14 |c_1|^2 \int_{s_{\rm min}}^{s_{\rm max}} 
\mbox{d}s 
\frac{\langle |\vec J\,|^2\rangle}{(s-M_{K_1})^2 + M_{K_1}\Gamma_{K_1}}~~.
\eea

The extraction of the remaining coefficients is
slightly complicated by the fact that two of the functions $f_i(x)$ are not 
linearly independent
\bea
f_6(x) = \frac12 f_1(x) - f_5(x)~,\quad
f_7(x) = \frac12(f_2(x) + f_4(x))~~.
\eea
Nevertheless, fitting the angular distribution $F(\tilde \theta)$ in 
(\ref{ftheta}) to a sum of five independent functions of $\tilde\theta$, allows 
one to determine the linear combination $d_2-d_4$. 
(Alternatively, $d_2-d_4$ may be obtained by projecting out the part 
proportional to $\cos\tilde \theta$ by integration with an appropriate weight 
function, $\phi(\tilde \theta) = 5\cos \tilde \theta - 7\cos^3 \tilde \theta$).
Since this combination is proportional to 
$\lambda_\gamma$, the photon polarization parameter can be extracted from the 
following ratio
\bea
\lambda_\gamma = \frac{d_2 - d_4}{4R_1 d_1 - 12R_2 d_3}~~.
\eea

\section{Calculations of $R_1$ and $R_2$}

The parameters $R_1$ and $R_2$ measure the up-down asymmetry of the photon 
momentum with respect to the $K\pi\pi$ decay plane in the $K_{\rm res}$ frame. 
For decays dominated by $K_1$ and $K^*_2$ they are defined such that 
$\langle \cos\tilde \theta\rangle = R_1\lambda_\gamma$ and 
$\langle \cos\tilde \theta\rangle = R_2\lambda_\gamma$, respectively.
In the case of $K_1$ the integrated up-down asymmetry is given by 
$(3/2)R_1\lambda_\gamma$,
while in the case of $K^*_2$ the integrated asymmetry vanishes.
Here we will show that $R_1$ and $R_2$ can be computed quite reliably with 
some dependence on the hadronic parameters describing $K_1$ and $K^*_2$ decays.

We start by considering the decays of the $K_1(1400)$ resonance discussed 
in Sec.~IV.1, and discuss first decays to final states involving a neutral
pion, $K^0\pi^+\pi^0$ and $K^+\pi^-\pi^0$.
The parameter $R_1$, defined in Eqs.~(\ref{Ri}), can be
calculated by applying Eqs.~(\ref{Ci}), (\ref{CK*rho}), (\ref{BW2}) and 
(\ref{JK}).
Its value depends somewhat on two pairs of hadronic parameters: $|c_D/c_S|$ and
$\delta_{D/S}$, the magnitude and phase of the ratio of $D$ to $S$-wave 
amplitudes in $K_1 \to K^*\pi$, and $\kappa_S$ and $\alpha_S$, the magnitude and 
phase of the ratio of S-wave amplitudes in $K_1 \to \rho K$ and $K_1 \to 
K^*\pi$. Varying these parameters in the measured ranges, $|c_D/c_S| = 
1.75 \pm 0.22~{\rm GeV}^{-2}, \delta_{D/S} = 260^\circ \pm 20^\circ,
\kappa_S = 0.32 \pm 0.32, 20^\circ \le \alpha_S \le 60^\circ$, we find values 
of $R_1$
in the range $R_1 = 0.22 \pm 0.03$, where in (\ref{Ri}) we integrate over the 
entire Dalitz plot. The uncertainty in $R_1$ combines an uncertainty of
$\pm 0.02$ due to the ratio of $D$ and $S$ wave amplitudes in the $K^*\pi$ 
channel, and an uncertainty of $\pm 0.02$ due to the $\rho K$ amplitude.
We conclude that the integrated up-down asymmetry 
originating in $K_1$ alone is quite large, $(0.33 \pm 0.05)\lambda_\gamma$.

The corresponding asymmetry in the $K^+\pi^+\pi^-$ and $K^0\pi^+\pi^-$
channels is smaller. Here only one $K^*\pi$ intermediate state 
contributes, and the dominant asymmetry is due to interference between $D$ and 
$S$ wave amplitudes in this channel. It is proportional to $\sin\delta_{D/S}$
and, since $\delta_{D/S}$ is not far from  $3\pi/2$, one finds $R_1 \approx 
-0.07\sin\delta_{D/S} \approx 0.07$. The correction from the $\rho K$ channel 
may change this value by about $50\%$, depending on the value of the strong phase 
$\alpha_S$.

Next, consider the decays of $K^*_2(1430)$ resonance discussed in Sec.~IV.2,
from which the value of $R_2$ defined in Eq.~(\ref{Ri}) is calculated through 
Eqs.~(\ref{BW2}) and (\ref{JK}).
This value depends on the magnitude $|\kappa|$ and phase
$\alpha$ of the ratio of $K^*_2 \to \rho K$ and $K^*_2 \to K^*\pi$
amplitudes, which we take as specified in Sec. IV.2, $|\kappa| = 2.38,~
-30^\circ \le \alpha \le 30^\circ$.  The value obtained when integrating over 
the entire Dalitz plot is rather small, varying as a function of $\alpha$ in the 
range $R_2 = 0.01 - 0.05$.

Larger values for $R_2$, which are less sensitive to corrections from $\rho K$,
are obtained by restricting the region in the Dalitz plot over which one 
integrates. To be specific, let us consider a square region
(A), $0.71 \mbox{GeV}^2 \leq s_{13}, s_{23} \leq 0.89$ GeV$^2$, centered at
the $K^*$ mass, $s_{13}=s_{23} = M_{K^*}^2$, of sides equal
to twice the $K^*$ width.
The value of $R_2$ in region (A) is dominated by the $K^*$ contributions and,
when neglecting the $\rho$ contribution, is given by
\beq\label{R_A}
R_{2A} = -\frac13\frac{\langle |\vec p_1\times \vec p_2|^3
\mbox{Im}\,\left (
B_{K^*}(s_{13})B^*_{K^*}(s_{23})\right )\mbox{sgn}(s_{13}-s_{23})\rangle_A}
{\langle |\vec p_1\times \vec p_2|^2 |\vec p_1 B_{K^*}(s_{23}) +
\vec p_2 B_{K^*}(s_{13})|^2\rangle_A} = 0.091~~.
\eeq
Including the $\rho$ contribution modifies this value only mildly to become
$R_A=0.071 \pm 0.002$, where we use the above values of $|\kappa|$
and $\alpha$.

One can easily see why the value of $R_{2A}$ is positive. While the variable in 
the denominator of (\ref{R_A}) is positive, the one in the numerator,
containing a factor $(s_{23}-s_{13})\mbox{sgn}(s_{13}-s_{23})$ is negative.
The quantity $\mbox{Im}\,\left ( B_{K^*}(s_{13})B^*_{K^*}(s_{23})\right )$
contains the relatively narrow $K^*$ width, $\Gamma_{K^*} = 51$ MeV.
One may imagine a higher excited $2^+$ resonance decaying to a pion and a 
wider $1^-$ $K$-resonance state, such as $K^*(1680)$, where subsequently 
$K^*(1680) \to K\pi$.
In this case a larger value of $R_2$ can be obtained due to a larger width.

\section{Feasibility of the method}

In order to estimate the number of $B$ mesons required for a feasible
measurement of the photon polarization parameter $\lambda_\gamma$
in $B\to K\pi\pi\gamma$, let us 
first assume for simplicity that one is able to measure separately decays 
through the $K_1(1400)$ $1^+$ state. This will require the least number of 
$B$'s, since this resonance state was shown to lead to much larger polarization 
effects than the other resonance states. For final states of the types
$K^0\pi^+\pi^0$ and $K^+\pi^-\pi^0$, we calculated the integrated up-down
asymmetry of the photon momentum with respect to the $K\pi\pi$ decay plane
and found it to be $(0.33 \pm 0.05)\lambda_\gamma$. In order to measure at 
three standard deviations an asymmetry of $-0.33$, as expected in the SM 
(where $\lambda_\gamma \approx -1$), one needs to observe a total of about 
80 charged and neutral $B$ and $\bar B$ decays to $K\pi\pi\gamma$ in these 
two channels.

When estimating the branching ratio for these events, we will assume 
that the exclusive $B$ decay branching ratio into $K_1(1400)\gamma$ is 
$0.7\times 10^{-5}$, as calculated in some models \cite{gK}. The decays 
$K_1 \to K\pi\pi$ are dominated by $K^*\pi$, ${\cal B}(K_1(1400) \to K^*\pi) = 
0.94 \pm 0.06$,
where we'll take the central value. Using isospin, one finds 
that $4/9$ of all $K^*\pi$ events in $K_1^+$ and $K_1^0$ decays occur in 
the two channels $K^0\pi^+\pi^0$ and $K^+\pi^-\pi^0$, respectively.
One must also include a factor $1/3$ for observing a $K_S$ (from $K^0$) 
through its $\pi^+\pi^-$ decays. Overall, we estimate an observable
branching ratio of ${\cal B} = 0.7\times 10^{-5}\times (4/9)0.94  
\simeq 0.3\times 10^{-5}$ into $(K^+\pi^-\pi^0)_{K_1(1400)}$ and 
${\cal B} \simeq 0.1\times 10^{-5}$ into $(K_S\pi^+\pi^0)_{K_1(1400)}$.
These branching ratios imply that, in order to observe the necessary 80
$K\pi\pi\gamma$ events and to measure their asymmetry at $3\sigma$, 
one needs at least $2\times 10^7 B\bar B$ pairs, including charged and
neutrals. 
This estimate does not include factors of efficiency and background,
which may increase the number of required $B$'s by an order of magnitude.

The $K\pi\pi$ invariant mass range $1300-1500$ MeV which we considered 
obtains also a contribution from the upper tail of a lower $1^+$ resonance at 
$1270$ MeV, which decays to $K^*\pi$ and $\rho K$ with branching ratios
of $16 \pm 5\%$  and $42 \pm 6\%$, respectively (see Table I). 
In order to suppress this contribution, which would interfere 
with the $K_1(1400)$ amplitude, one may study the upper half mass 
range, $m(K\pi\pi) = 1400-1500$ MeV, and extend it to 1600 MeV for higher 
statistics. This range, which includes about half of all 
$B\to K_1(1400)\gamma$ decays, may obtain a small
contribution from the lower tail of the wide $1^-~K^*(1680)$ resonance  
at 1717 MeV, decaying to $K^*\pi$ and $\rho K$ with branching ratios
of about $30\%$ each. This resonance does not interfere, however, with of 
the $K_1(1400)$ state whose strong polarization effect one is using to 
measure $\lambda_\gamma$.  

In the above estimate of the required number of $B$ mesons we have assumed 
separation of events originating from 
the $K_1(1400)$ resonance, rather than basing our estimate on the calculated
decay distribution which combines the three overlapping resonances. 
It would be interesting 
to study the efficiency of the method described in 
Sec.~V, which extracts $\lambda_\gamma$ from the decay distribution
combining all resonances. This challenging task is beyond the scope of this 
paper, and should be treated more professionally by experimental methods
when more data become available.

\section{Conclusions}

We studied a method for measuring in $B\to K\pi\pi\gamma$ the photon 
polarization parameter $\lambda_\gamma$ occurring in the effective weak 
Hamiltonian describing radiative $b$ quark decays.
The SM predicts that $\lambda_\gamma \approx -1~(+1)$ 
for $B^- (B^+)$ and $\bar B^0 (B^0)$ decays.
Different values, possibly with an opposite sign, can be obtained in extensions 
of the SM, such as the Left-Right model and Minimal Supersymmetry. 
The parameter $\lambda_\gamma$ was shown to be measured through an up-down 
asymmetry 
of the photon direction relative to the $K\pi\pi$ decay plane in the $K\pi\pi$ 
center of mass frame. In the SM the photon prefers to move in the hemisphere
of $\vec p_{\rm slow}\times \vec p_{\rm fast}$ in $B^-$ and $\bar B^0$ decays,
and in the opposite hemisphere in $B^+$ and $B^0$ decays.

We studied the amplitudes of $B\to K\pi\pi\gamma$ in the $K$ resonance region
for a few distinct charged modes. Combining contributions from several 
overlapping resonances
in a mass range near 1400 MeV, $K_1(1400),~K^*_2(1430)$ and $K^*(1410)$,
the general decay distribution was calculated. A method was proposed for using 
this distribution to determine the photon polarization parameter.
Based on an up-down asymmetry of $(0.33 \pm 0.05)\lambda_\gamma$ from 
$K_1(1400)$ alone, we conclude that a first 
measurement of $\lambda_\gamma$ can be performed with about $10^8$ $B\bar B$ 
pairs, combining charged and neutrals. This study can be performed at currently 
operating $B$-factories.

\medskip
We thank Yuval Grossman and Anders Ryd for an enjoyable collaboration 
which initiated this study.
This work was supported in part by the Israel Science Foundation
founded by the Israel Academy of Sciences and Humanities,
by the U. S. -- Israel Binational Science Foundation through Grant
No.\ 98-00237 and by the DOE under grant DOE-FG03-97ER40546.

\def \ajp#1#2#3{Am.\ J. Phys.\ {\bf#1}, #2 (#3)}
\def \apny#1#2#3{Ann.\ Phys.\ (N.Y.) {\bf#1}, #2 (#3)}
\def \app#1#2#3{Acta Phys.\ Polonica {\bf#1}, #2 (#3)}
\def \arnps#1#2#3{Ann.\ Rev.\ Nucl.\ Part.\ Sci.\ {\bf#1} (#3) #2}
\def \art{and references therein}
\def \cmts#1#2#3{Comments on Nucl.\ Part.\ Phys.\ {\bf#1} (#3) #2}
\def \cn{Collaboration}
\def \cp89{{\it CP Violation,} edited by C. Jarlskog (World Scientific,
Singapore, 1989)}
\def \efi{Enrico Fermi Institute Report No.\ }
\def \epjc#1#2#3{Eur.\ Phys.\ J. C {\bf#1}, #2 (#3)}
\def \f79{{\it Proceedings of the 1979 International Symposium on Lepton and
Photon Interactions at High Energies,} Fermilab, August 23-29, 1979, ed. by
T. B. W. Kirk and H. D. I. Abarbanel (Fermi National Accelerator Laboratory,
Batavia, IL, 1979}
\def \hb87{{\it Proceeding of the 1987 International Symposium on Lepton and
Photon Interactions at High Energies,} Hamburg, 1987, ed. by W. Bartel
and R. R\"uckl (Nucl.\ Phys.\ B, Proc.\ Suppl., vol.\ 3) (North-Holland,
Amsterdam, 1988)}
\def \ib{{\it ibid.}~}
\def \ibj#1#2#3{~{\bf#1}, #2 (#3)}
\def \ichep72{{\it Proceedings of the XVI International Conference on High
Energy Physics}, Chicago and Batavia, Illinois, Sept. 6 -- 13, 1972,
edited by J. D. Jackson, A. Roberts, and R. Donaldson (Fermilab, Batavia,
IL, 1972)}
\def \ijmpa#1#2#3{Int.\ J.\ Mod.\ Phys.\ A {\bf#1}, #2 (#3)}
\def \ite{{\it et al.}}
\def \jhep#1#2#3{JHEP {\bf#1}, #2 (#3)}
\def \jpb#1#2#3{J.\ Phys.\ B {\bf#1} (#3) #2}
\def \jpg#1#2#3{J.\ Phys.\ G {\bf#1}, #2 (#3)}
\def \lg{{\it Proceedings of the XIXth International Symposium on
Lepton and Photon Interactions,} Stanford, California, August 9--14 1999,
edited by J. Jaros and M. Peskin (World Scientific, Singapore, 2000)}
\def \lkl87{{\it Selected Topics in Electroweak Interactions} (Proceedings of
the Second Lake Louise Institute on New Frontiers in Particle Physics, 15 --
21 February, 1987), edited by J. M. Cameron \ite~(World Scientific, Singapore,
1987)}
\def \kdvs#1#2#3{{Kong.\ Danske Vid.\ Selsk., Matt-fys.\ Medd.} {\bf #1},
No.\ #2 (#3)}
\def \ky85{{\it Proceedings of the International Symposium on Lepton and
Photon Interactions at High Energy,} Kyoto, Aug.~19-24, 1985, edited by M.
Konuma and K. Takahashi (Kyoto Univ., Kyoto, 1985)}
\def \mpla#1#2#3{Mod.\ Phys.\ Lett.\ A {\bf#1} (#3) #2}
\def \nat#1#2#3{Nature {\bf#1} (#3) #2}
\def \nc#1#2#3{Nuovo Cim.\ {\bf#1} (#3) #2}
\def \nima#1#2#3{Nucl.\ Instr.\ Meth. A {\bf#1} (#3) #2}
\def \npb#1#2#3{Nucl.\ Phys.\ B~{\bf#1}, #2 (#3)}
\def \os{XXX International Conference on High Energy Physics, 27 July
-- 2 August 2000, Osaka, Japan}
\def \PDG{Particle Data Group, D. E. Groom \ite, \epjc{15}{1}{2000}}
\def \pisma#1#2#3#4{Pis'ma Zh.\ Eksp.\ Teor.\ Fiz.\ {\bf#1} (#3) #2 [JETP
Lett.\ {\bf#1} (#3) #4]}
\def \pl#1#2#3{Phys.\ Lett.\ {\bf#1} (#3) #2}
\def \pla#1#2#3{Phys.\ Lett.\ A {\bf#1} (#3) #2}
\def \plb#1#2#3{Phys.\ Lett.\ B {\bf#1}, #2 (#3)}
\def \pr#1#2#3{Phys.\ Rev.\ {\bf#1} (#3) #2}
\def \prc#1#2#3{Phys.\ Rev.\ C {\bf#1} (#3) #2}
\def \prd#1#2#3{Phys.\ Rev.\ D {\bf#1}, #2 (#3)}
\def \prl#1#2#3{Phys.\ Rev.\ Lett.\ {\bf#1}, #2 (#3)}
\def \prp#1#2#3{Phys.\ Rep.\ {\bf#1} (#3) #2}
\def \ptp#1#2#3{Prog.\ Theor.\ Phys.\ {\bf#1} (#3) #2}
\def \rmp#1#2#3{Rev.\ Mod.\ Phys.\ {\bf#1} (#3) #2}
\def \rp#1{~~~~~\ldots\ldots{\rm rp~}{#1}~~~~~}
\def \si90{25th International Conference on High Energy Physics, Singapore,
Aug. 2-8, 1990}
\def \slc87{{\it Proceedings of the Salt Lake City Meeting} (Division of
Particles and Fields, American Physical Society, Salt Lake City, Utah, 1987),
ed. by C. DeTar and J. S. Ball (World Scientific, Singapore, 1987)}
\def \slac89{{\it Proceedings of the XIVth International Symposium on
Lepton and Photon Interactions,} Stanford, California, 1989, edited by M.
Riordan (World Scientific, Singapore, 1990)}
\def \smass82{{\it Proceedings of the 1982 DPF Summer Study on Elementary
Particle Physics and Future Facilities}, Snowmass, Colorado, edited by R.
Donaldson, R. Gustafson, and F. Paige (World Scientific, Singapore, 1982)}
\def \smass90{{\it Research Directions for the Decade} (Proceedings of the
1990 Summer Study on High Energy Physics, June 25--July 13, Snowmass,
Colorado),
edited by E. L. Berger (World Scientific, Singapore, 1992)}
\def \tasi{{\it Testing the Standard Model} (Proceedings of the 1990
Theoretical Advanced Study Institute in Elementary Particle Physics, Boulder,
Colorado, 3--27 June, 1990), edited by M. Cveti\v{c} and P. Langacker
(World Scientific, Singapore, 1991)}
\def \yaf#1#2#3#4{Yad.\ Fiz.\ {\bf#1} (#3) #2 [Sov.\ J.\ Nucl.\ Phys.\
{\bf #1} (#3) #4]}
\def \zhetf#1#2#3#4#5#6{Zh.\ Eksp.\ Teor.\ Fiz.\ {\bf #1} (#3) #2 [Sov.\
Phys.\ - JETP {\bf #4} (#6) #5]}
\def \zpc#1#2#3{Zeit.\ Phys.\ C {\bf#1}, #2 (#3)}
\def \zpd#1#2#3{Zeit.\ Phys.\ D {\bf#1} (#3) #2}

\end{document}